# SPEECIAL RELATIVITY DESCRIPTION OF THE HEAT PROPAGATION IN MINKOWSKI SPACETIME

J. Marciak – Kozłowska*  and  M.Kozłowski**

*Institute of Electron Technology, Al. Lotników 32/46

02 – 668 Warsaw, Poland

Corresponding author: e-mail miroslawkozlowski@aster.pl

**Abstract**

In this paper we investigate the heat transport induced by continuous laser beams up to an intensity of about $10^{29}$ Watt/cm$^2$. We maintain that up to this intensity nonlinear effects are negligible and that the application of the linear hyperbolic heat transport equation is fully justifiable. We show that the Fourier diffusion equation gives the speed of diffusion, $v > c$ and breaks the causality of the thermal processes in Minkowski space-time. For hyperbolic heat transport v<c and causality is valid

## 1  INTRODUCTION

In the interaction of the laser beam with a thin solid target, fast ion beams and electrons are generated in profusion. When the energy of the laser beam is sufficiently high to achieve relativistic speeds of the ions and electrons then the transport of thermal energy must be described by relativistic equations.

In this paper we develop a description of heat transport in Minkowski space-time. Within the context of special relativity theory we investigate the Fourier and hyperbolic diffusion equations. We calculate the speed of thermal diffusion in the Fourier approximation and show that for high energy laser beams the thermal diffusion exceeds the speed of light. We show that this result breaks the causality of thermal phenomena in Minkowski space-time. The same phenomena we describe within the framework of the hyperbolic thermal diffusion equation and show that in that case the speed of diffusion is always less than the speed of light.

In this paper we discuss the applicability of the linear hyperbolic diffusion equation in describing the ultra high energy laser beam interaction with matter.

We show that non-linear effects are negligible up to an intensity of about $10^{29}$ Watt/cm$^2$.

## 2 MINKOWSKI SPACE-TIME

We may use the concept that the speed of light *in vacuo* is the upper limit of speed and also in which a signal can travel between two events to establish whether or not any two events could be connected. In the interest of simplicity we shall work with one space dimension $x_1 = x$ and the time dimension $x_o = ct$ of the Minkowski space-time. Let us consider events (1) and (2): their Minkowski interval *Δs* satisfies the relationship:

$$\Delta s^2 = c^2 \Delta t^2 - \Delta x^2 \qquad (1)$$

Without loss of generality we can take Event 1 to be at $x = 0$, $t = 0$. Then Event 2 can be only related to Event 1 if it is possible for a signal travelling at the speed of light, to connect them. If Event 2 is at (*Δx, cΔt*), its relationship to Event 1 depends on whether *Δs* > 0, = 0, or < 0.

We may summarise the three possibilities as follows:

Case A  **time-like** interval, $|\Delta x_A| < c\Delta t$, or $\Delta s^2 > 0$. Event 2 can be related to Event 1,

    Events 1 and 2 can be in causal relation.

Case B  **light-like** interval, $|\Delta x_B| = c\Delta t$, or $\Delta s^2 = 0$. Event 2 can only be related to Event 1 by a light signal.

Case C  s**pace-like** interval $|\Delta x_A| > c\Delta t$, or $\Delta s^2 < 0$. Event 2 cannot be related to Event 1, for in that case $v > c$.

Now let us consider case C in more detail. At first sight, it seems that in case C we can find out the reference frame in which two Events $c^>$ and $c^<$ always fulfils the relationship $t_{c^>} - t_{c^<}$ > 0. but this is not true. If we choose the inertial frame U' in which $t_{c^>} - t_{c^<}$ > 0 and the reference frame U is moving with a speed *V* relative to U', with

$$V = c\frac{c(t'_{c_>} - t'_{c_<})}{x'_{c_<} - x'_{c_>}} \qquad (2)$$

then for a speed $V<c$

$$\left|\frac{c(t'_{c_>} - t'_{c_<})}{x'_{c_<} - x'_{c_>}}\right| < 1 \qquad (3)$$

Let us calculate $t_{c_>} - t_{c_<}$ in the reference frame U

$$t_{c_>} - t_{c_<} = \frac{1}{\sqrt{1-\frac{V^2}{c^2}}}\left[\frac{V}{c^2}(x'_{c_>} - x'_{c_<}) + (t'_{c_>} - t'_{c_<})\right] =$$

$$\frac{1}{\sqrt{1-\frac{V^2}{c^2}}}\left[\frac{t'_{c_>} - t'_{c_<}}{x'_{c_>} - x'_{c_<}}(x'_{c_>} - x'_{c_<}) + (t'_{c_>} - t'_{c_<})\right] = 0 \qquad (4)$$

For a higher $V$ we have $t_{c_>} - t_{c_<} < 0$. This implies that for space-like intervals the sign of $t_{c_>} - t_{c_<}$ depends on the speed $V$, i.e. the causality relation for space-like events is not valid.

## 3  FOURIER DIFFUSION EQUATION AND SPECIAL RELATIVITY

In paper [1] the speed of diffusion signals was calculated

$$v = \sqrt{2D\omega} \qquad (5)$$

where

$$D = \frac{\hbar}{m} \qquad (6)$$

and $\omega$ is the angular frequency of the laser pulses. Considering equations (5) and (6) one obtains

$$v = c\sqrt{2\frac{\hbar\omega}{mc^2}} \qquad (7)$$

and $v \geq c$ for $\hbar\omega \geq mc^2$.

From equation (7) we conclude that for $\hbar\omega > mc^2$ the Fourier diffusion equation is in contradiction with the special relativity theory and thus breaks the causality in transport phenomena.

## 4  HYPERBOLIC DIFFUSION AND SPECIAL RELATIVITY

In monograph [2] the hyperbolic model of the thermal transport phenomena was formulated. It was shown that the description of the ultra-short thermal energy transport needs the hyperbolic diffusion equation (one dimensional transport)

$$\tau \frac{\partial^2 T}{\partial t^2} + \frac{\partial T}{\partial t} = D \frac{\partial^2 T}{\partial x^2} \tag{9}$$

In equation (9) $\tau = \dfrac{\hbar}{m\alpha^2 c^2}$ is the relaxation time, $m$ = mass of the thermal carrier, $\alpha$ is the coupling constant and $c$ is the speed of light in vacuum, $T(x,t)$ is the temperature field and $D = \hbar/m$.

In paper [1] the speed of thermal propagation $v$ was calculated

$$v = \frac{2\hbar}{m} \sqrt{-\frac{m}{2\hbar}\tau\omega^2 + \frac{m\omega}{2\hbar}(1+\tau^2\omega^2)^{1/2}} \tag{10}$$

Considering that $\tau = \hbar/m\alpha^2 c^2$ equation (10) can be written as

$$v = \frac{2\hbar}{m} \sqrt{-\frac{m}{2\hbar}\frac{\hbar\omega^2}{mc^2\alpha^2} + \frac{m\omega}{2\hbar}(1+\frac{\hbar^2\omega^2}{m^2 c^4 \alpha^4})^{1/2}} \tag{11}$$

For

$$\frac{\hbar\omega}{mc^2\alpha^2} < 1, \quad \frac{\hbar\omega^2}{mc^2} < 1 \tag{12}$$

one obtains from equation (11)

$$v = \sqrt{\frac{2\hbar}{m}\omega} \tag{13}$$

Formally equation (13) is the same as equation (7) but considering the inequality (11) we obtain

$$v = \sqrt{\frac{2\hbar\omega}{m}} = \sqrt{2}\alpha c < c$$

(14)

and the causality is not broken.

For

$$\frac{\hbar\omega}{mc^2} > 1; \qquad \frac{\hbar\omega}{\alpha^2 mc^2} > 1 \qquad (15)$$

we get from equation (11)

$$v = \alpha c, \qquad v < c \qquad (16)$$

Considering equations (14) and (16) we conclude that the hyperbolic diffusion equation (9) describes the thermal phenomena in accordance with special relativity theory and the causality is not broken irrespective of laser beam energy.

When the amplitude of the laser beam approaches the critical electric field of quantum electrodynamics (Schwinger field [3]) the vacuum becomes polarised and electron – positron pairs are created in vacuum [3]. At a distance equal to the Compton length, $\lambda_C = \hbar/m_e c$, the work of the critical field on an electron is equal to the electron rest mass energy $m_e c^2$, i.e. $eE_{Sch}\lambda_C = m_e c^2$. The dimensionless parameter

$$\frac{E}{E_{Sch}} = \frac{e\hbar E}{m_e^2 c^3} \qquad (17)$$

becomes equal to unity for an electromagnetic wave intensity of the order of

$$I = \frac{c}{r_e \lambda_C^2} \frac{m_e c^2}{4\pi} \cong 4.7 \cdot 10^{29} \tfrac{W}{cm^2} \qquad (18)$$

where $r_e$ is the classical electron radius [4]. For such ultra high intensities the effects of nonlinear quantum electrodynamics plays a key role: laser beams excite virtual electron – positron pairs. As a result the vacuum acquires a finite electric and magnetic susceptibility which lead to the scattering of light by light. The cross section for the photon – photon interaction is given by:

$$\sigma_{\gamma\gamma\to\gamma\gamma} = \frac{973}{10125}\frac{\alpha^3}{\pi^2} r_e^2 \left(\frac{\hbar\omega}{m_e c^2}\right)^6, \qquad (19)$$

for $\hbar\omega/m_e c^2 < 1$ and reaches its maximum, $\sigma_{max} \approx 10^{-20} cm^2$ for $\hbar\omega \approx m_e c^2$ [3].

Considering equations (18) and (19) we conclude that the linear hyperbolic diffusion equation is valid only for laser intensities $I \leq 10^{29}$ W/cm². At higher intensities the nonlinear hyperbolic diffusion equation must be formulated and solved.

# Table 1

## Hierarchical structure of the thermal excitation

| Interaction | $\alpha$ | $mc^2\alpha$ |
|---|---|---|
| Electromagnetic | $137^{-1}$ | $0.5/137$ |
| Strong | $\dfrac{15}{100}$ | $\dfrac{140 \cdot 15}{100}$ for pions |
| | | $\dfrac{1000 \cdot 15}{100}$ for nucleons |
| Quark - Quark | 1 | $417^*$ |

\* D.H. Perkins, Introduction to high energy physics, Addison – Wesley, USA 1987